\documentclass{article}

\usepackage{amsfonts}
\usepackage{epsfig}

\begin{document}

\title{\textsc{Sestieri of Venice}}

\vspace{1cm}

\author{ D. Volchenkov and Ph. Blanchard
\vspace{0.5cm}\\
{\it  BiBoS, University Bielefeld, Postfach 100131,}\\
{\it D-33501, Bielefeld, Germany} \\
{\it Phone: +49 (0)521 / 106-2972 } \\
{\it Fax: +49 (0)521 / 106-6455 } \\
{\it E-Mail: VOLCHENK@Physik.Uni-Bielefeld.DE}}
\large

\date{\today}
\maketitle

\begin{abstract}

We have investigated space syntax of Venice by means of random
walks. Random walks being defined on an undirected graph establish
the Euclidean space in which distances and angles between nodes
acquire the clear statistical interpretation. The properties of
nodes with respect to random walks allow partitioning the city
canal network into disjoint divisions which may be identified with
the traditional divisions of the city (sestieri).
\end{abstract}

\vspace{0.5cm}

\leftline{PACS: 89.65.Lm, 89.75.Fb, 05.40.Fb, 02.10.Ox }
 \vspace{0.5cm}

\leftline{ Keywords: Complex networks, city space syntax }

\section{The Sestieri of Venice}
\label{sec:sestieri_of_Venice}
\noindent

Spectral methods
can be implemented  in order to visualize graphs of not very large
 multi-component networks \cite{Volchenkov2007}.
City districts constructed
accordingly to different development principles
 in different historical epochs can be  envisioned on
the dual graph representation of space syntax.

We investigate the segmentation of the spatial network of
96 canals in Venice (that stretches across 122 small
 islands between which the canals serve the function of roads) in
accordance to its historical divisions.
\begin{figure}[ht]
 \noindent
\begin{center}
\epsfig{file=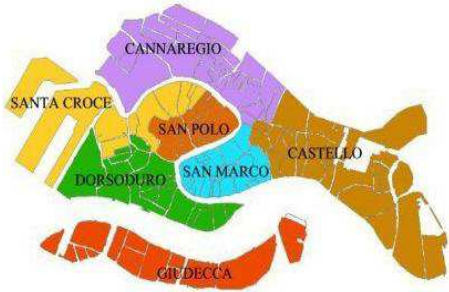,  angle= 0,width =6cm, height =4cm}
  \end{center}
\caption{\small The sestieri are the primary traditional divisions of Venice.
The image has been taken from 'Portale di Venezia' at http://www.guestinvenice.com/}
\label{Fig2_05}
\end{figure}
The sestieri are the primary traditional divisions of Venice (see Fig.~\ref{Fig2_05}):
 Cannaregio, San Polo, Dorsoduro, Santa Croce, San Marco and Castello, Giudecca.
The oldest settlements in Venice had appeared from the 6$^\mathrm{th}$ century in Dorsoduro,
along the Giudecca Canal. By the 11$^\mathrm{th}$century, settlement had spread across to
the Grand Canal. The Giudecca island is composed of 8 islets separated by canals
dredged in the 9$^\mathrm{th}$ century  when the area was divided among the rebelling nobles.
San Polo is the smallest of the six sestieri of Venice, covering just 35 hectares
along the Grand Canal. It is one of the oldest parts of the city, having been
settled before the 9$^\mathrm{th}$ century, when it and San Marco (lying in the heart
 of the city) formed part of the Realtine Islands.
  Cannaregio named after the Cannaregio Canal is the second largest district of
   the city. It was developed from the 11$^\mathrm{th}$ century. Santa Croce occupies
    the north west part of the main islands lying on land only created form
    the late Middle ages to the twentieth century.
The district Castello  grew up from the 13$^\mathrm{th}$ century.

In the present paper, we address the following question:
Given a spatial network of a city, is it possible to uncover its historical
and functional divisions directly from its space syntax?

In Sec.~\ref{sec:Graphrepresentations}, we discuss the primary and
dual graph representations of urban environments. The dual graph
representation has been extensively studied in space syntax theory
which is instrumental in predicting human behavior in urban
environments. In Sec.~\ref{sec:Traffic_equilibrium}, we
demonstrate that  space syntax is related to the traffic
equilibrium state of a transport network, and Markov's transition
operators naturally appear in the space syntax context embedding
city space syntax into Euclidean space. We build the dual graph
representation of Venetian canals in
Sec.~\ref{sec:Dual_Graph},~\ref{sec:Graph-Partitioning} and then
perform the Principal Component Analysis of Venetian space syntax
in Sec~\ref{sec:PCA}. The properties of nodes with respect to
random walks allow partitioning the city canal network into
disjoint divisions which may be identified with the traditional
divisions of the city (sestieri).

\section{Graphs and Space Syntax of Urban Environments}
\label{sec:Graphrepresentations}
\noindent

Urban space is of rather large scale
 to be seen from a single viewpoint;
maps provide us with its representations by
means of abstract symbols facilitating our
 perceiving and understanding of a city.
 The middle scale and small scale
maps are usually based on Euclidean geometry
 providing spatial objects with precise coordinates
 along their edges and outlines.

There is a long tradition of research articulating urban
environment form using graph-theoretic
principles originating from the paper of Leonard Euler
(see \cite{GraphTheory}).
Graphs have long been regarded as the basic structures
for
representing forms where topological relations are firmly
 embedded within Euclidean
space.
The widespread use of graph theoretic analysis in geographic
 science had been reviewed in \cite{Haggett}
establishing it as central to spatial analysis of urban environments.
In \cite{Kansky}, the basic graph
theory methods had been applied  to the measurements of
 transportation networks.

Network analysis has long been a basic function of geographic
information systems
(GIS) for a variety of applications, in which computational
 modelling of an urban network is based on a graph view in
 which the intersections of linear features are regarded as
nodes, and connections between pairs of nodes are represented
as edges \cite{MillerShaw}.
Similarly, urban forms are usually represented as the patterns
of identifiable urban elements such as
locations or areas (forming nodes in a graph) whose relationships
to one another are often associated with linear
transport routes such as streets within cities \cite{Batty}.
Such planar graph representations define locations or points in
Euclidean plane as nodes or vertices $\{ i\}$, $i=1,\ldots, N$, and
 the edges linking  them together as $i\sim j$, in
 which $\{i,j\}=1,2,\ldots,N.$ The value of a link can
 either be binary, with the value $1$ as $i\sim j$, and
 $0$ otherwise, or be equal to actual physical distance
 between nodes,  $\mathrm{dist}(i,j)$, or to some weight $w_{ij}>0$ quantifying
a certain characteristic property of the link.
We shall call a planar graph representing the Euclidean space
embedding of an urban network as its {\it primary graph}.
 Once a spatial system has been identified and
represented by a graph in this way, it can be subjected to
the graph theoretic analysis.

A {\it spatial network} of a city is a network of the spatial
 elements of urban environments. They
are derived from maps of {\it open spaces} (streets, places, and roundabouts).
Open spaces may be broken down into components; most simply, these
might be street segments, which can be linked into a network via
their intersections and analyzed as a networks of {\it movement
choices}. The study of spatial configuration is instrumental in
predicting {\it human behavior}, for instance, pedestrian
movements in urban environments \cite{Hillier96}. A set of
theories and techniques for the analysis of spatial configurations
is called {\it space syntax} \cite{Jiang98}. Space syntax is
established on a quite sophisticated speculation that the
evolution of built form can be explained in analogy to the way
biological forms unravel \cite{SSyntax}. It has been developed as
a method for analyzing space in an urban environment capturing its
quality
 as being comprehendible and easily navigable \cite{Hillier96}.
Although,  in its initial form, space syntax was focused mainly on
patterns of pedestrian movement in cities, later the  various
space syntax measures of urban configuration had been found to be
correlated with the different aspects of social life,
\cite{Ratti2004}.

Decomposition of a space map into a complete set of
intersecting axial lines,  the fewest and
longest lines of sight that pass through every open space comprising any system,
produces an axial map or an overlapping convex map respectively.
Axial lines and convex spaces may be treated as the {\it spatial elements}
 (nodes of a morphological graph),
 while either the {\it junctions} of axial lines or the {\it overlaps} of
 convex  spaces may be considered as the edges linking  spatial elements
 into a single  graph unveiling the
topological relationships  between all open elements of the urban space.
In what follows,  we shall call this morphological representation of urban network
as a {\it dual graph}.

The transition to a dual graph is a topologically non-trivial
transformation of a planar primary graph into a non-planar one which
encapsulates the hierarchy and structure of the urban area and also corresponds
 to perception of space that people experience when
travelling along routes through the environment.

In Fig.~1, we have presented the glossary establishing a correspondence
 between several typical elements of urban environments and the certain subgraphs
 of dual graphs.
The dual transformation replaces the 1D open segments (streets) by the
 zero-dimensional nodes, Fig.~1(1).
\begin{figure}[ht]
\label{F1_11}
 \noindent
\begin{center}
\begin{tabular}{llrr}
 1. &\epsfig{file=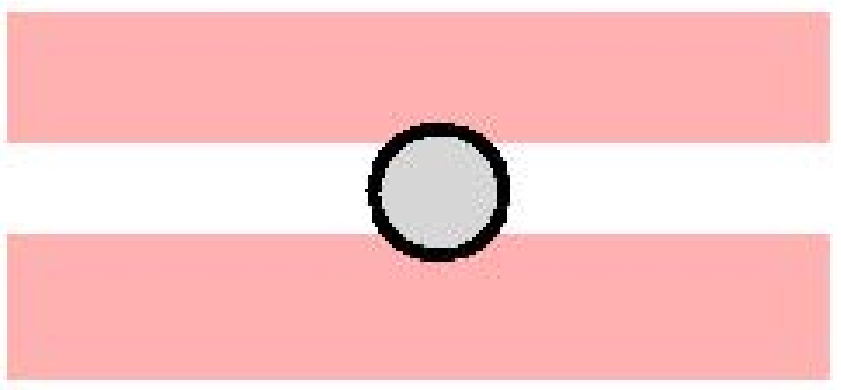, width=3.0cm, height =1.5cm}&2.&
 \epsfig{file=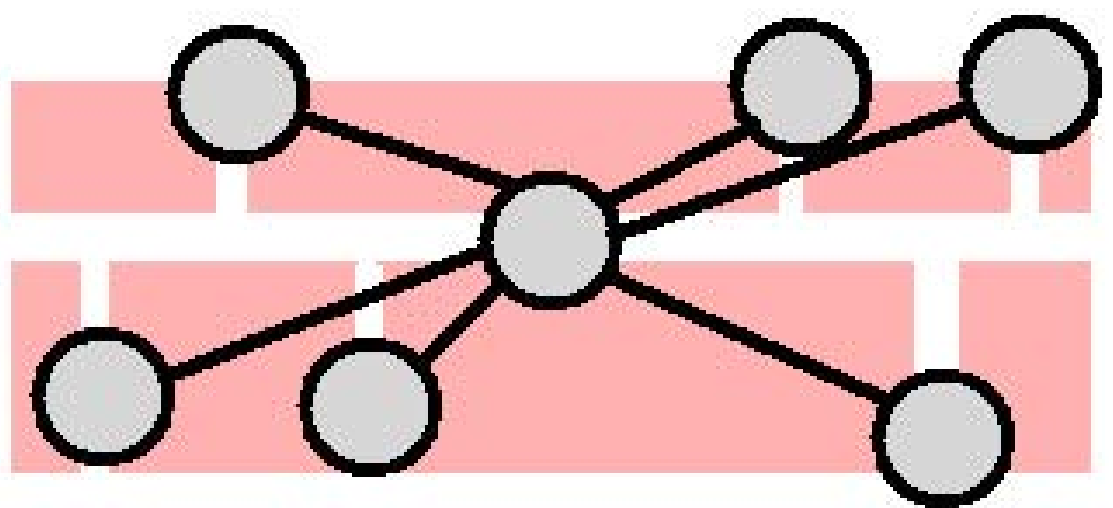, width=3.0cm, height =1.5cm} \\
 3. &\epsfig{file=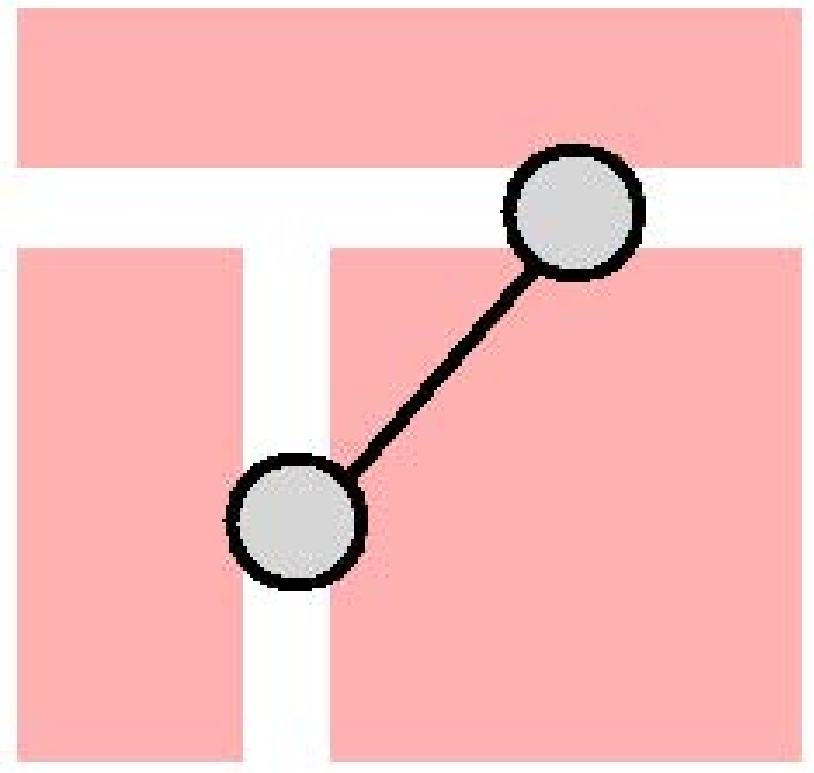,width=3.0cm, height =3.0cm}&4.&
 \epsfig{file=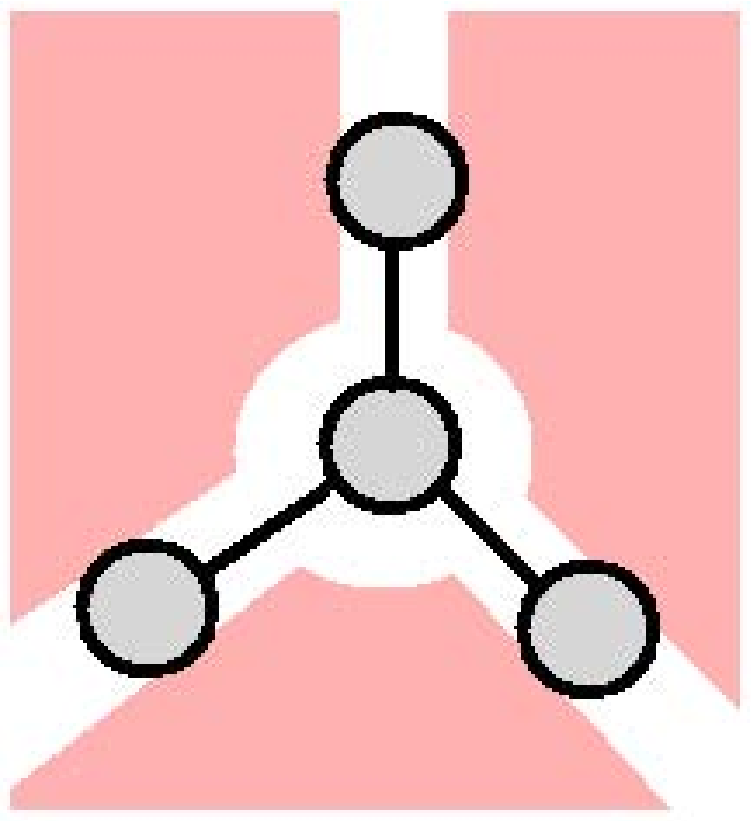, width=3.0cm, height =3.0cm} \\
 5. &\epsfig{file=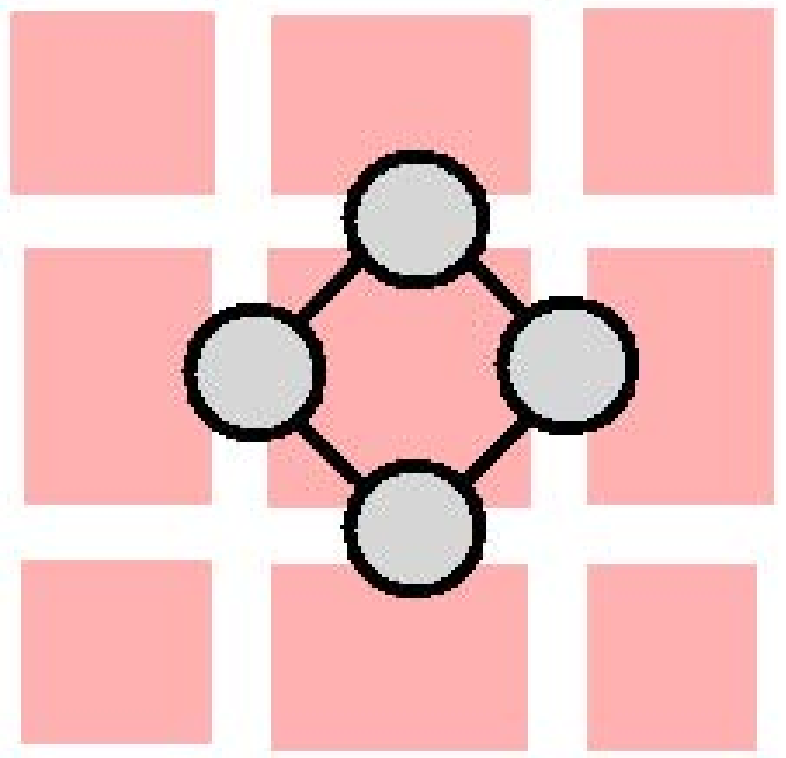,width=3.0cm, height =3.0cm}&6.&
 \epsfig{file=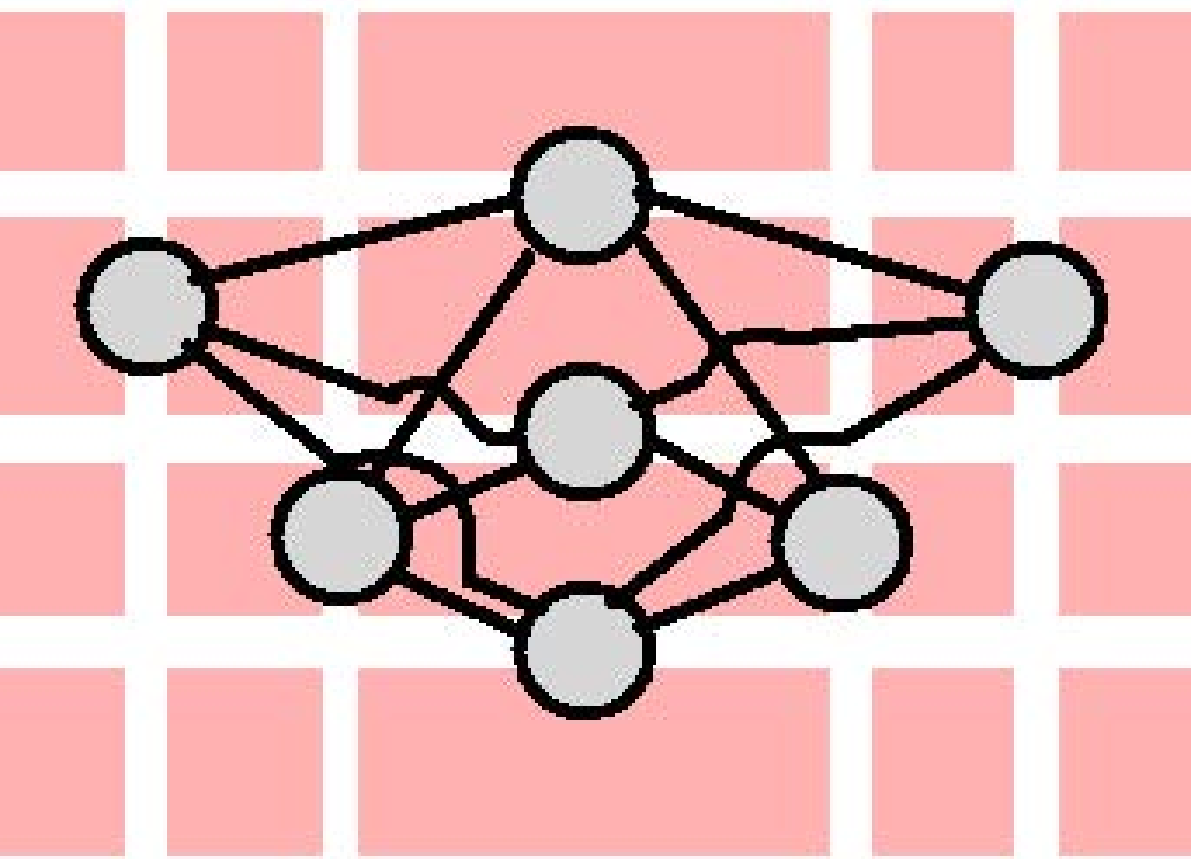, width=3.0cm, height =3.0cm} \\
\end{tabular}
\end{center}
\caption{The dual transformation glossary between the typical elements of
urban environments and the certain subgraphs of dual graphs.}
\end{figure}

The sprawl like developments consisting of a number of blind passes branching
 off a main route are changed to the star subgraphs having a hub and a number
 of client nodes, Fig.~1(2). Junctions and crossroads are replaced with
 edges connecting the corresponding nodes of the dual graph, Fig.1(3).
  Places and roundabouts are considered as the independent topological
  objects and acquire the individual IDs being nodes in the dual
  graph Fig.~1(4). Cycles are converted into  cycles of
  the same lengthes Fig.~1(5). A regular grid pattern
  is shown in Fig.~1(6). Its dual graph representation
  is called a {complete bipartite} graph, where the set of vertices
  can be divided into two disjoint subsets  such that no edge has
   both end-points in the same subset, and every line joining the
   two subsets is present, \cite{Krueger89}.
These sets can be naturally interpreted as those of the vertical
and horizontal edges in the primary graphs (streets and avenues).
In bipartite graphs, all closed paths are of even length, \cite{Skiena}.

 It is the dual
graph transformation which allows to separate the effects of order
and of structure while analyzing a transport network on the
morphological ground. It converts the repeating geometrical
elements expressing the order in the urban developments into the
{\it twins nodes}, the pairs of nodes such that any other is
adjacent either to them both or to neither of them. Examples
of twins nodes can be found in Fig.~1(2,4,5,6).

\section{Traffic Equilibrium, Space Syntax, and Random Walks}
\label{sec:Traffic_equilibrium}
\noindent

The concept of {\it equilibrium}, the condition of a system in which all
 competing influences are balanced, is a key theoretical element in any
  branch of science. The notion of traffic equilibrium had been
  introduced by J.G. Wardrop \cite{Wardrop:1952} and then generalized
  by \cite{Beckmann:1956}  to a fundamental concept of {\it network
  equilibrium}  with many potential applications such as the establishing
  of rigorous mathematical foundations for the analysis of congested
  transport networks.
Wardrop's traffic equilibrium  \cite{Wardrop:1952} is strongly tied to
city space syntax since it is required that while attaining the equilibrium
all travellers have enough knowledge of the transport network they use.
 Because of the complexity of traffic situation in the network, the route
 choice decisions taken by travellers are  not always objectively optimal.
However, there is another link between the traffic equilibrium and space
syntax which has never been discussed in the literature.

Given a connected undirected graph $G(V,E)$, in which $V$ is the
set of nodes and $E$ is the set of edges, we can define the traffic
 volume $f: E\to(0,\infty[$ through every edge $e\in E$. It then follows
  from the Perron-Frobenius theorem that a linear equation
\begin{equation}
\label{Lim_equilibrium}
f(e)\,=\, \sum_{e'\,\in\, E}\,f(e')\,\exp\left(\,-h\,\ell\left(e'\right)\,\right)
\end{equation}
has a unique positive solution $f(e)>0$, for every edge $e\in E$, for a
fixed positive constant $h>0$ and a chosen set of positive  {\it metric
 length} distances $\ell(e)>0$. This solution is naturally identified
 with the traffic equilibrium state of the transport network defined on
  $G$, in which the permeability of edges depends upon their lengths.
The parameter $h$  is called the volume entropy of the graph $G$, while
 the volume of $G$ is defined as the sum
\begin{equation}
\label{volume}
\mathrm{Vol}(G)=\frac 12\sum_{e\in E}\ell(e).
\end{equation}
The degree of a node $v\in V$ is the number of its neighbors in
$G$, $\deg(v)=k_v$. It has been shown in \cite{Lim:2005} that
among all undirected connected graphs of normalized volume,
$\mathrm{Vol}(G)=1$, which are not cycles and $k_v\ne 1$ for all
nodes,
 the minimal possible value of the volume entropy,
\begin{equation}
\label{min_entropy}
\min(h)=\frac 12\sum_{v\in V}k_v\,\log\left(k_v-1\right)
\end{equation}
 is attained
for the length distances
\begin{equation}
\label{ell_min}
\ell(e)\,=\,\frac {\log\left(\left(k_{i(e)}-1\right)
\left(k_{t(e)}-1\right)\right)}{2\,\min(h)},
\end{equation}
where $i(e)\in V$ and $t(e)\in V$ are the initial and terminal vertices
of the edge $e\in E$ respectively. It is then obvious that (\ref{ell_min})
and $\min(h)$ being substituted into (\ref{Lim_equilibrium}) change the
operator $\exp\left(-h \ell(e)\right)$ to a symmetric Markov transition operator,
\begin{equation}
\label{Markov_transition}
f(e)\,=\, \sum_{e'\,\in\, E}\,\frac{f(e')}{\sqrt{\left(k_{i(e)}-1\right)
\left(k_{t(e)}-1\right)}},
\end{equation}
which rather describes time reversible random walks over edges than over
nodes. The flows satisfying (\ref{Lim_equilibrium}) with the Markov operator
(\ref{Markov_transition}) meet the mass conservation property,
\begin{equation}
\label{mass_conserv}
\sum_{i\sim j}f_{ij}=\pi_j,\quad \sum_{j\in V}\pi_j=1,
\end{equation}
for some node constants $\pi_i>0$. Other solutions $f(e)>0$
obtained for $h\,> \min(h)$ describe equilibrium flows with
termination of travellers. The Eq.(\ref{Markov_transition})
unveils the indispensable role Markov's
 chains defined on edges play in equilibrium traffic modelling and exposes
 the degrees of nodes as a key determinant of the transport networks properties.

Random walks embed connected undirected graphs
into Euclidean space, in which distances and angles acquire the
clear statistical interpretation.

Any graph representation  naturally arises as an outcome of categorization,
when we abstract a real world system by eliminating all but one of its features
 and by the grouping of things (or places) sharing a common attribute by
 classes or categories. For instance, the common attribute of all open spaces
 in city space syntax is that we can move through them. All elements called
 nodes that fall into one and the same group $V$ are considered essentially
 identical; permutations of them within the  group are of no consequence.
The symmetric group $\mathbb{S}_{N}$ consisting of all permutations of $N$
elements
($N$ is the cardinality of the set $V$) constitute the symmetry group of $V$.
If we denote by $E\subseteq V\times V$ the set of ordered pairs of nodes called
edges, then  a graph is a map $G(V,E): E \to K\subseteq\mathbb{R}_{\,+}$
(we suppose that the graph has no multiple edges).

The nodes of $G(V,E)$ may be weighted with respect to some  {\it  measure}
 $m=\sum_{i\in V} m_i \delta_i,$ specified by a set of positive numbers $m_i> 0$.
  The space $\ell^2(m)$ of square-assumable functions with respect to the
   measure $m$ is the {\it Hilbert space} $\mathcal{H}$ (a complete inner
   product  space). Among all linear operators defined on $\mathcal{H}$,
   those  {\it invariant} under the permutations of nodes are of essential
   interest since they reflect the symmetry of the graph. Although there are
   infinitely many such operators, only those which maintain  conservation
   of a quantity may describe a physical process. The Markov transition
   operators which share the property of  {\it probability conservation}
   considered in theory of random walks on graphs are among them.
Laplace operators describing diffusions on graphs meet the {\it mean value}
 property ({\it mass conservation}); they give another example \cite{Smola2003}
  studied in spectral graph theory.

 Markov's operators on Hilbert space  form the natural language of complex networks theory.
Being defined on connected undirected graphs, a Markov transition operator $T$ has a unique
{\it equilibrium} state $\pi$ (a stationary distribution of random walks)
 such that
\begin{equation}
\label{stationary}
\pi\, T\,=\,\pi, \quad \pi\,=\,\lim_{t\to\infty}\,\sigma\, T^{\,t},
\end{equation}
for any density $\sigma\in \mathcal{H}$ ($\sigma_i\geq 0$,
$\sum_{i\in V} \sigma_i=1$). There is a unique measure $m_\pi
=\sum_{i\,\in\, V} \pi_i\delta_i $ related to the stationary
distribution $\pi$ with respect to which the Markov operator $T$
is {\it self-adjoint},
\begin{equation}
\label{s_a_analogue}
\widehat{T}=\,\frac 12
\left( \pi^{1/2}\,\, T\,\,
\pi^{-1/2}+\pi^{-1/2}\,\, T^\top\,\,
 \pi^{1/2}\right),
\end{equation}
where $T^\top$ is the adjoint operator. The orthonormal ordered set of real
 eigenvectors $\psi_i$, $i=1\ldots N$, of the symmetric operator $\widehat{T}$
  establishes the {\it basis} in  $\mathcal{H}$. In quantitative theory of
  random walks  defined on graphs \cite{Lovasz:1993,Aldous} and in spectral
  graph theory \cite{Chung:1997}, the properties of graphs are studied in
   relationship to the  eigenvalues and eigenvectors of self-adjoint operators
   defined on them. In particular, the symmetric transition operator defined on
   undirected graphs is
\begin{equation}
\label{T_symm}
\widehat{T_{ij}}\,=\,\left\{
\begin{array}{ll}
\frac 1{\sqrt{k_i\,k_j}}, & i\sim j\\
 0, & \mathrm{otherwise}.
\end{array}
\right.
\end{equation}
Its first eigenvector $\psi_1$ belonging to the largest eigenvalue $\mu_1=1$,
\begin{equation}
\label{psi_1}
\psi_1
\,\widehat{ T}\, =\,
\psi_1,
\quad \psi_{1,i}^2\,=\,\pi_i,
\end{equation}
describes the {\it local} property of nodes (connectivity), $\pi_i=k_i/2M,$
 where $2M=\sum_{i\in V} k_i$, while the remaining eigenvectors
 $\left\{\,\psi_s\,\right\}_{s=2}^N$ belonging to the eigenvalues
  $1>\mu_2\geq\ldots\mu_N\geq -1$ delineate the {\it global} connectedness of the graph.

Markov's symmetric transition operator $\widehat{T}$  defines a {\it projection}
 of any density $\sigma\in \mathcal{H}$ on the eigenvector $\psi_1$ of the
  stationary distribution $\pi$,
\begin{equation}
\label{project}
\sigma\,\widehat{T}\,
=\,\psi_1 + \sigma^\bot\,\widehat{T},\quad \sigma^\bot\,=\,\sigma-\psi_1.
\end{equation}
Thus, it is clear that any two densities $\sigma,\rho\,\in\,\mathcal{H}$ differ
 with respect to random walks only by their dynamical components,
\[
(\sigma-\rho)\,
 \widehat{T}^t\,=\,(\sigma^\bot -\rho^\bot)\,
\widehat{T}^t,
\]
for all $t\,>\,0$.
Therefore, we can define a
distance between any two densities which they acquire  with respect to random walks by
\begin{equation}
\label{distance}
\left\|\,\sigma-\rho\,\right\|^2_T\, =
\, \sum_{t\,\geq\, 0}\, \left\langle\, \sigma-\rho\,\left|T^t
\right|\, \sigma-\rho\,\right\rangle.
\end{equation}
 or, in the spectral form,
\begin{equation}
\label{spectral_dist}
\begin{array}{lcl}
\left\|\,\sigma-\rho\,\right\|^2_T & = & \sum_{t\,\geq 0}\, \sum_{s=2}^N\,
 \mu^t_s \,\left\langle\,
\sigma-\rho\,|\,\psi_s\right\rangle\!\left\langle\, \psi_s
\,|\, \sigma-\rho\,\right\rangle \\
 & = &  \sum_{s=2}^N\,\frac{\left\langle\, \sigma-\rho\,|\,
 \psi_s\right\rangle\!\left\langle\, \psi_s
\,|\, \sigma-\rho\,\right\rangle}{\,1\,-\,\mu_s\,},
\end{array}
\end{equation}
where we have used  Dirac's  bra-ket notations especially
convenient in working with inner products and
rank-one operators in Hilbert space.

If we introduce the new inner product in $\mathcal{H}(V)$ by
\begin{equation}
\label{inner-product}
\left(\,\sigma,\rho\,\right)_{T}
\,= \, \sum_{t\,\geq\, 0}\, \sum_{s=2}^N
\,\frac{\,\left\langle\, \sigma\,|\,\psi_s\,\right\rangle\!
\left\langle\,\psi_s\,|\,\rho \right\rangle}{\,1\,-\,\mu_s\,}
\end{equation}
for all  $\sigma,\rho\,\in\, \mathcal{H}(V),$
then (\ref{spectral_dist}) can be written as
\begin{equation}
\label{spectr-dist2}
\left\|\,\sigma-\rho\,\right\|^2_T\, =
\left\|\,\sigma\,\right\|^2_T +
\left\|\,\rho\,\right\|^2_T  -
2 \left(\,\sigma,\rho\,\right)_T,
\end{equation}
 in which
\begin{equation}
\label{sqaured_norm}
\left\|\, \sigma\,\right\|^2_T\,=\,
\,\sum_{s=2}^N \,\frac{\left\langle\, \sigma\,|\,\psi_s\,\right\rangle\!
\left\langle\,\psi_s\,|\,\sigma\, \right\rangle}{\,1\,-\,\mu_s\,}
\end{equation}
is the squared norm of  $\sigma\,\in\, \mathcal{H}(V)$ with respect to
random walks.
We accomplish the description of the $(N-1)$-dimensional Euclidean
space structure associated to random walks by mentioning that
given two densities $\sigma,\rho\,\in\, \mathcal{H}(V),$ the
angle between them can be introduced in the standard way,
\begin{equation}
\label{angle}
\cos \,\angle \left(\rho,\sigma\right)=
\frac{\,\left(\,\sigma,\rho\,\right)_T\,}
{\left\|\,\sigma\,\right\|_T\,\left\|\,\rho\,\right\|_T}.
\end{equation}
Random walks embed connected undirected graphs into Euclidean
space that can be used in order to compare
 nodes
and to retrace
 the optimal coarse-graining
representations.
Namely,  let us consider
the density $\delta_i$ which equals 1 at
the node $i\,\in\, V$ and zero for all other nodes. It takes form
$\upsilon_i\,=\,\pi^{-1/2}_i\,\delta_i$ with respect to the measure
$m_\pi$. Then, the squared norm of  $\upsilon_i$ is given by
\begin{equation}
\label{norm_node}
\left\|\,\upsilon_i\,\right\|_T^2\, =\,{\frac 1{\pi_i}\,\sum_{s=2}^N\,
\frac{\,\psi^2_{s,i}\,}{\,1-\mu_s\,}},
\end{equation}
where $\psi_{s,i}$ is the $i^{\mathrm{th}}$-component of the
eigenvector $\psi_s$. In quantitative theory of random walks \cite{Lovasz:1993},
the quantity (\ref{norm_node}) is known as the {\it access time} to a target node
quantifying the expected number
of  steps
required for a random walker
to reach the node
$i\in V$ starting from an
arbitrary
node  chosen randomly
among all other
nodes  with respect to
the stationary distribution $\pi$.

The notion of spatial
segregation acquires a statistical interpretation
with respect to random walks defined on the graph. In urban
spatial networks encoded by their dual graphs, the access times
 $\left\|\,\upsilon_i\,\right\|_T^2$
vary strongly from one open space
to another: the norm of a street that can be easily
 reached (just in a few random syntactic steps)
from
any other street in the city
 is minimal,
while it could be very large for a
statistically segregated
street.

The Euclidean distance between any two nodes of the graph $G$
established by random walks,
\begin{equation}
\label{commute}
\begin{array}{lcl}
K_{i,j} &=& \left\|\, \upsilon_i-\upsilon_j\,\right\|^2_T \\
 & =& \sum_{s=2}^N\,\frac 1{1-\mu_s}\,\left(\frac{\psi_{s,i}}
 {\sqrt{\pi_i}}-\frac{\psi_{s,j}}{\sqrt{\pi_j}}\right)^2
\end{array}
\end{equation}
is known as {\it commute time}
in  quantitative
theory of random walks
and  equals to
the expected number of steps
required for a random walker
starting at $i\,\in\, V$ to visit
$j\,\in\, V$ and
then to return to
$i$ again,  \cite{Lovasz:1993}.

It is important to mention that
the cosine of an angle calculated in accordance to
 (\ref{angle}) has the structure of
Pearson's coefficient of linear correlations
 that reveals it's natural
statistical interpretation.
Correlation properties of flows
of random walkers
passing by different paths
 have been remained beyond the scope of
previous  studies devoted to complex
networks and random walks on graphs.
The notion of angle between any two nodes in the
graph arises naturally as soon as we
become interested in
the strength and direction of
a linear relationship between
two random variables,
the flows of random walks moving through them.
If the cosine of an angle (\ref{angle}) is 1
(zero angles),
there is an increasing linear relationship
between the flows of random walks through both nodes.
Otherwise, if it is close to -1 ($\pi$ angle),
  there is
a decreasing linear relationship.
The  correlation is 0 ($\pi/2$ angle)
if the variables are linearly independent.
It is important to mention that
 as usual the correlation between nodes
does not necessary imply a direct causal
relationship (an immediate connection)
between them.

\section{Dual Graph of Venetian Canals}
\label{sec:Dual_Graph}
\noindent

While analyzing the canal network of Venice,
we have assigned an identification number to each of 96 city canals.
Then the dual graph representation for the canal network is
constructed by mapping canals encoded by the same ID
into nodes of the dual graph and intersections among each
pair of canals
 into edges connecting the
corresponding nodes.
The problem of segmentation is closely related to the problem of three
dimensional (3D) visual representations.

In order to obtain the 3D visual representation of the dual graph
for the canal network of Venice, we use the spectral properties of
symmetric transition operator  (\ref{T_symm}).

 The $(x_i,y_i,z_i)$ coordinates of the $i^\mathrm{th}$-node of the dual graph
in 3D space are given by the relevant $i^\mathrm{th}$-components of three
eigenvectors taken from the ordered set  $\{\psi_k\}$, $k = 2\,\ldots N$.
Possible segmentations and symmetries of dual
graphs can be discovered visually by using different triples
of eigenvectors if the number of nodes in the graph is not very large.
\begin{figure}[ht]
 \noindent
\begin{center}
\epsfig{file=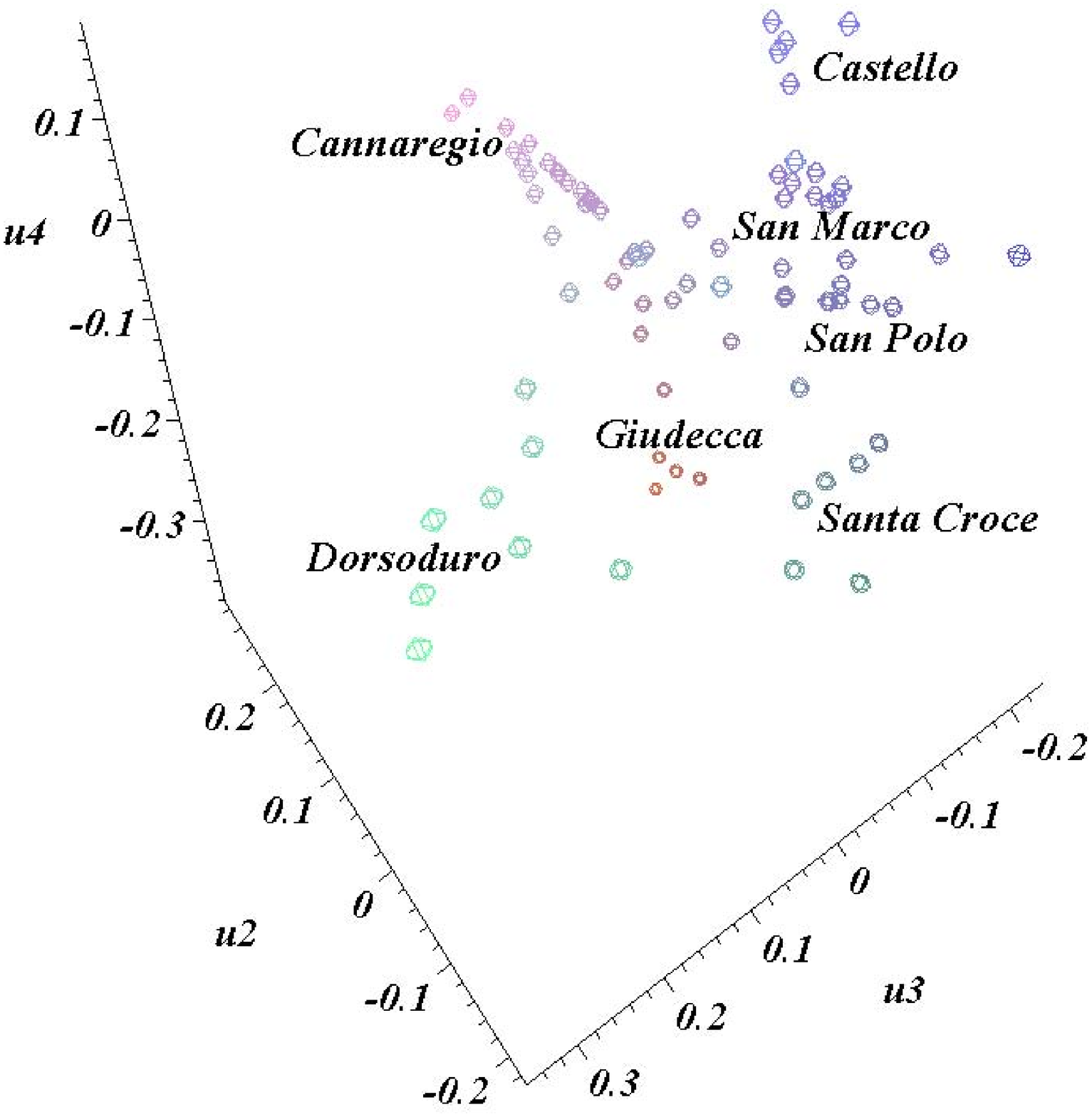, angle= 0,width =7cm, height =7cm}
  \end{center}
\caption{\small The segmentation of the dual graph of Venetian canals using
 three eigenvectors $\left[\psi_2,\psi_3,\psi_4\right]$.
The nodes of dual graph can be partitioned
  into classes which can be almost precisely identified with the historical
   divisions of Venice. }
\label{Fig2_06}
\end{figure}
In Fig.~\ref{Fig2_06}, we have presented the results of segmentation for
the dual graph of Venetian canals using the eigenvectors
$\left[\,\psi_2,\psi_3,\psi_4\,\right]$
belonging to the primary eigenvalues  $1\,>\,\mu_2\,>\,\mu_3\,>\mu_4$.
Nodes of the dual graph belonging to one and the same
 city district  developed in a certain historical epoch
are located on one and the same quasi-surface in the Euclidean
space established by random walks.

Primary eigenvectors of Markov's transition operator defined
on the dual graph representation  of a network
indicate the directions
 in which the equilibrium flows have maximal "extensions".
The use of these
  eigenvectors as a basis helps us to divide the nodes of the dual graph into
   classes which can be almost precisely identified with the historical
    city districts. Let us note that the implementation of other
     eigenvectors as the basis for the 3D representations of the dual graph
       worsens the quality of segmentation in a sense that
it turns to be incompatible with the traditional sestieri of Venice.
The slowest modes of diffusion process described by the primary eigenvectors
 allow detecting city modules of different accessibility.

Due to the proper normalization, the components of eigenvectors
  play the role of the
Participation Ratios (PR) which quantify the effective numbers of
nodes participating in a given eigenvector with a significant
weight. This characteristic has been used in \cite{ESMS} and by
other authors to describe the modularity of complex networks.
However, PR is not
a well defined quantity in the case of eigenvalue multiplicity
since the different vectors in the
eigenspace corresponding to the degenerate mode would obviously
have different PR.

\section{Graph Partitioning by Random Walks}
\label{sec:Graph-Partitioning}
\noindent

Visual segmentation of networks based on 3D representations of
 their dual graphs is not always feasible.
Furthermore, the result of such a segmentation may essentially
depend on which eigenvectors
have been chosen as the basis for the 3D representation. The
computation of eigenvectors for large matrices can be
time and resource consuming, and therefore it is important to have
a good estimation on the minimal number of eigenvectors required
for the proper graph segmentation.

The graph partitioning problem seeks to partition a weighted undirected
graph $G$ into $n$ weakly connected components $\Gamma_1,\ldots\Gamma_n$ such that
 $\bigcup_{i=1}^n\Gamma_n\subset G$ and either their properties share some common trait or
the graphs nodes belonging to them are close to each other according to some distance
 measure defined on nodes of the graph.
A number of different graph partitioning strategies for undirected weighted  graphs
 have been studied in  connection with Object Recognition and Learning in
  Computer Vision \cite{Vision}.

In Ec.~\ref{sec:Traffic_equilibrium}, we have shown
that random walks being
introduced on a connected undirected graph $G$
establish the $(N-1)$-dimensional Euclidean space
in which every pair of nodes, $i$ and $j$, appear at some distance
\begin{equation}
\label{dist_RW}
\mathrm{dist}_T(i,j)\,=\,\sqrt{K_{i,j}}
\end{equation}
 where $K_{i,j}$ is the commute time (\ref{commute}) of random walks between $i$ and $j$.
The random walks distance (\ref{dist_RW}) can be used as a measure of similarity between
 any two nodes in $G$.
Namely, every node $i\in V$ of an undirected graph $G$
may be represented by a vector
 ${\bf z}^{(i)}\,\in\,\mathbb{R}^{\,(N-1)}$ in the
$(N-1)$-dimensional Euclidean space associated to random walks,
\begin{equation}
\label{node_vector}
{\bf z}^{(i)}\,=\,\left( \frac {\psi_{2,i}}
{\sqrt{\pi_i\,\left(1-\mu_2\right)}},\ldots,\frac {\psi_{N,i}}
{\sqrt{\pi_i\,\left(1-\mu_N\right)}}\right).
\end{equation}
We then assign each vector (\ref{node_vector}) to one of $n$ clusters,
 $\Gamma_l$ whose center ({\it centroid}),
\begin{equation}
\label{centroid}
{\bf m}^{(l)}\,=\,
\frac 1{\left|\Gamma_l\right|}\,
\sum_{j=1}^{\left|\Gamma_l\right|}\, {\bf z}^{(j)},
\end{equation}
  is the nearest to ${\bf z}^{(i)}$ with respect to the
  distance (\ref{dist_RW}).
The objective we try to achieve is to minimize the total intra-cluster
 variance of the resulting partition $\mathcal{P}$ of the graph $G$
 into $n$ clusters, the squared error function (s.e.f.),
\begin{equation}
\label{sef1}
\mathrm{sef}
(\mathcal{P})\,=\,\sum_{l=1}^n\,\sum_{s=1}^{\left|\Gamma_l\right|}
 \,\left|\,z_{s}^{(l)}-{\bf m}^{(l)}\,\right|^2.
\end{equation}
If we denote the $(N-1)$-dimensional unity vector by ${\bf e}=(1,1,\ldots,1)^\top$
and ${\bf Z}\,=\,\left[{\bf z}^{(1)},\ldots{\bf z}^{(N)}\right]$ is
the $N\times (N-1)$-matrix of coordinates the node acquire in
the Euclidean space associated to random walks, then it is clear that
\begin{equation}
\label{sef2}
\begin{array}{lcl}
\mathrm{sef}\left(\mathcal{P}\right) &= &
\sum_{l=1}^n \, \left|\,{\bf Z}-{\bf m}^{(l)}{\bf e}^\top\,\right|^2\\
 &=& \sum_{l=1}^n \, \left|\,{\bf Z}- {\bf P}_l\,\right|^2,
\end{array}
\end{equation}
where
$
{\bf P}_l \,=\, \left({\bf 1}_{\Gamma_l}\,-\,\frac{{\bf e\,e}^\top}
{\left|\Gamma_l\right|}\right)
$
 is the projection operator of nodes onto the cluster $\Gamma_l$.
 Since ${\bf P}_l^2\,=\,{\bf P}_l$, we immediately obtain that
\begin{equation}
\begin{array}{ll}
\mathrm{sef}
\left(\mathcal{P}\right) &
=\sum_{l=1}^n \, \mathrm{tr}\,\left({\bf Z}^\top{\bf P}_l{\bf Z}\right) \\
& =
\mathrm{tr}\,\left({\bf Z}^\top{\bf Z}\right)\,-\,\mathrm{tr}\,\left({\bf X}^\top{\bf Z}^\top\,{\bf Z}{\bf X}\,\right),
\end{array}
\label{sef3}
\end{equation}
in which
\begin{equation}
\label{norm_indicators}
{\bf X}=
\left[\frac{\chi^1}{\sqrt{\left|\Gamma_1\right|}},\ldots,\frac{\chi^{\,n}}{\sqrt{\left|\Gamma_n\right|}}\right]
\end{equation}
 is the rectangular orthogonal $n\times N$-matrix
 $\left({\bf X}^\top{\bf X}\,=\,{\bf 1}\right)$
 of the normalized indicator
 vectors
\begin{equation}
\label{parition_indicator}
\chi^l_i\,=\,
\left\{
\begin{array}{ll}
1, & i\,\in \,\Gamma_l,\\
0, & i \,\notin \,\Gamma_l.
\end{array}
\right.
\end{equation}
Considering  elements of the
 ${\bf Z}^\top{\bf Z}$ matrix as measuring similarity between nodes, we
 can show following \cite{Zha1} that the Euclidean distance (\ref{dist_RW})
leads to Euclidean inner-product similarity which can be replaced
by a general Mercer kernel \cite{Saitoh,Wahba} uniquely
represented by a positive semi-definite matrix $K_{i,j}$.

If we then
relax the discrete structure of $\bf X$
by assuming that $\bf X$ is
 an arbitrary orthonormal matrix, the minimization of the objective function
$\mathrm{sef}(\mathcal{P})$ is reduced to the trace
maximization problem,
\begin{equation}
\label{maxim_trace}
\max_{{\bf X}^\top{\bf X}\,=\,{\bf 1}_{\,N-1}}\,\,
\mathrm{tr}\left(\,{\bf X}^\top\,{\bf Z}^\top{\bf Z}\,{\bf X}\,\right).
\end{equation}
A standard result in linear algebra (proven by K.Fan in 1949 \cite{Fan})
 provides a global
solution to the trace optimization problem:
Given a symmetric matrix ${\bf S}$ with eigenvalues
 $\lambda_1\geq\ldots\geq\lambda_n\geq\ldots\geq \lambda_N$, and the matrix
  of corresponding eigenvectors, $\left[\,{\bf u}_1,\ldots, {\bf u}_N\,\right]$, the
   maximum of $\mathrm{tr}\,\left(\,{\bf Q}^\top{\bf S}{\bf Q}\,\right)$ over all
   $n$-dimensional orthonormal matrices $\bf Q$ such that
    ${\bf Q}^\top{\bf Q}\,=\,{\bf 1}_n$ is given by
\begin{equation}
\label{maximiz}
\max_{{\bf Q}^\top{\bf Q}\,=\,{\bf 1}_n}\,\,\mathrm{tr}\,
\left(\,{\bf Q}^\top{\bf S}{\bf Q}\,\right)\,=\,
\sum_{k=1}^n \,\lambda_k,
\end{equation}
and the optimal $n$-dimensional orthonormal matrix
\begin{equation}
\label{Q_maixim}
{\bf Q}\,=
\, \left[\,{\bf u}_1,\ldots, {\bf u}_n\,\right]{\bf R}
\end{equation}
where ${\bf R}$ is an arbitrary orthogonal $n\times n$ matrix
(describing a rotation transformation in $\mathbb{R}^n$).

The result (\ref{maximiz}-\ref{Q_maixim}) relates the problem of
network segmentations  to the investigation of  $n$  primary
eigenvectors of a symmetric matrix defined on the graph nodes,
 \cite{Golub,Dhilon_ACM}. The eigenvectors ${\bf u}_{\,i\,>\,1}$
 have both positive and negative entries, so that in general
the matrix
 $\left[\,{\bf u}_1,\ldots, {\bf u}_n\,\right]$
differs substantially from
 that one comprising of the discrete cluster indicator vectors which have strictly
 positive entries.

It is important to note that  even for not very large $n$ it
may be rather difficult to compute the appropriate  $n\,\times\, n$
orthonormal transformation matrix $\bf R$ which recovers the necessary
 discrete cluster indicator structures.
 Furthermore, it can be shown that
the postprocessing of eigenvectors into the cluster indicator vectors can be
 reduced to an optimization problem with $n(n-1)/2\,-\,1$ parameters \cite{Ding}.
 Several methods have been proposed to obtain the partitions from
the eigenvectors of various similarity matrices
 (see \cite{Dhillon},\cite{Bach_Jordan} for a review).
 In the next section, we use the ideas of Principal Component
  Analysis (PCA) in order to bypass the orthonormal transformation.

\section{Principal Component Analysis of Venetian Canals}
\label{sec:PCA}
\noindent

In statistics, Principal Component Analysis (PCA) is used for the reducing size
 of a data set.
It is achieved by the optimal linear transformation retaining
the subspace that has largest variance (a lower-order principal component)
 and ignoring higher-order ones \cite{PCA,PCA2}.

Given an operator $S$ self-adjoint with respect
to  the  measure $m$
defined on a connected undirected graph $G$,
it is well known
  that the eigenvectors of the symmetric matrix ${\bf S}$
form an ordered orthonormal basis $\left\{\,\phi_k\,\right\}$
  with real eigenvalues
$\mu _1\geq\ldots\geq \mu_N$.
The ordered orthogonal basis
 represents the directions of the
 variances of variables
described by $S$.

If we
 consider the Laplace operator, $L\,=\,1-T,$ defined on $G$,
its  eigenvalues can be interpreted as
the inverse characteristic time scales
of the diffusion
  process such that the smallest eigenvalues
correspond to the stationary distribution
 together with the slowest diffusion
   modes involving the most
significant amounts of flowing commodity.
Therefore, while describing a network by means of the
Laplace operator,
we must  arrange the eigenvalues in increasing order,
$\lambda _1\leq\ldots\leq\lambda_n\leq\ldots\leq \lambda_N$, and
examine the ordered orthogonal basis of eigenvectors,
 $\left[\,{\bf f}_1,\ldots {\bf f}_N\,\right]$.

 The number of
components which
may be detected in a network
with regard to
a certain dynamical process defined on that
depends upon the number of {\it essential
eigenvectors} of the relevant self-adjoint operator.
There is
 a simple time scale argument
which we use in order to determine the number of
applicable eigenvectors.

It is obvious that while observing the network
 close to an equilibrium state
during short time, we detect flows resulting from
 a large number of transient  processes
evolving toward the stationary  distribution
and being characterized by the relaxation times
 $\propto\lambda^{-1}_k$.
While measuring the flows in
sufficiently long time $\tau$,
we may discover just $n$ different
eigenmodes, such that
\begin{equation}
\label{tau_lambda}
\lambda_1\,<\,\ldots\,\leq\,\lambda_n\,\leq\,\tau^{\,-1}\,<\,\ldots\,\leq
\,\lambda_N.
\end{equation}
In general, the longer is the time of measurements $\tau$,
the less is the
number of eigenvectors we
have to take into account in network component
 analysis of the network.
Should the time of measurements  is fixed, we can
determine the number  of required eigenvectors.

In the what-following, we consider the symmetric ("normalized")
Laplace operator,  \cite{Chung:1997},
\begin{equation}
\label{norm_Lapalce}
\widehat{ L_{ij}}\,=\,\delta_{ij}-\widehat{T_{ij}},
\end{equation}
where $\widehat{T_{ij}}$ is the symmetric Markov transition operator (\ref{T_symm}).

\subsection{Low dimensional representations of transport  networks
by the principal directions}

In order to obtain the best quality segmentation,
it is convenient to center the $n$ primary eigenvectors.
The {\it centroid} vector
(representing the center of mass of
the set $\left[\,{\bf f}_1,\ldots {\bf f}_n\,\right]$)
is calculated as
the arithmetic mean,
\begin{equation}
\label{centroid_f}
{\bf m}\,=\,\frac 1n \,\sum_{k=1}^{n}\, {\bf f}_{k}.
\end{equation}
Let us denote the $n\,\times\, N$ matrix of $n$
centered eigenvectors  by
\[
{\bf F}\,=\,\left[\,{\bf f}_1\,-\,{\bf m},\ldots \,{\bf f}_N\,-\,{\bf m}\right].
\]
Then, the symmetric matrix of {\it covariances}
 between the entries of eigenvectors $\{{\bf f}_k\}$
is the
product of ${\bf F}$ and its adjoint ${\bf F}^\top$,
\begin{equation}
\label{Covariance}
\mathrm{\bf Cov}\,=\,\frac {\,\,{\bf F}\,{\bf F}^\top\,}{N\,-\,1}
\end{equation}
It is important to note that the
correspondent
 Gram matrix ${\bf F}^\top\,{\bf F}\,/(N-1) \,\equiv\, {\bf 1}$
due to
the orthogonality of the basis eigenvectors.
\begin{figure}[ht]
 \noindent
\begin{center}
\epsfig{file=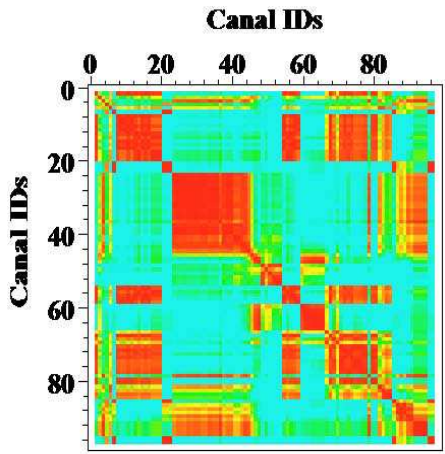, angle= 0,width =6cm, height =6cm}
  \end{center}
\caption{\small The correlation matrix calculated for the dual graph
representation of Venetian canals calculated for the system of
the first 7 eigenvectors of
the normalized Laplace operator (\ref{norm_Lapalce}). The
entries  of the  matrix are ranked from
1 (red) to -1 (blue).}
\label{Fig2_07}
\end{figure}
The main contributions in
the
symmetric matrix $\mathrm{\bf Cov}$ are related
to the groups of nodes
\begin{equation}
\label{directions}
\mathrm{\bf Cov}{\ }{\bf u}_k\,=\,\sigma_k\,{\bf u}_k,
\end{equation}
 which can be identified by means of the
 eigenvectors
$\{\,{\bf u}_k\,\}$ associated to the first largest
eigenvalues among
 $\sigma_1\,\geq\, \sigma_2,$ $\dots,\,\geq \,\sigma_N$.
By ordering the eigenvectors  in decreasing order
 (largest first), we create
an ordered orthogonal basis with the
  first eigenvector having the direction of largest variance of the
   components of  $n$ eigenvectors $\{{\bf f}_k\}$.
Let us note that
    due to the structure of ${\bf F}$ only the first $n-1$
    eigenvalues $\sigma_k$ are not trivial.
In accordance to the standard PCA notation,
the eigenvectors of the covariance
matrix ${\bf u}_k$  are called the
{\it principal directions} of the network
with respect to the diffusion process defined by
the operator $S$. A low dimensional representation of the network
is given by its principal directions
$\left[\,{\bf u}_1,\ldots,
{\bf u}_{n-1}\,\right],$ for $n\,<\,N$.

Diagonal
elements of the matrix $\mathrm{\bf Cov}$
quantify
the component variances of the eigenvectors
$\left[\,{\bf f}_1,\ldots {\bf f}_n\,\right]$
around their mean
 values (\ref{centroid_f}) and may be ample
essentially for large networks. Therefore,
 it is practical for us
to use
the standardized {\it correlation} matrix,
\begin{equation}
\label{Corr}
{\mathrm Corr}_{ij}\,=\,\frac{\mathrm{Cov}_{ij}}{\sqrt{\mathrm{Cov}_{ii}}
\sqrt{\mathrm{Cov}_{jj}}},
\end{equation}
instead of the covariance matrix
${\mathrm{\bf Cov}}$.
It is important to note that
the diagonal elements of (\ref{Corr})
equal 1, while the off-diagonal
elements are the Pearson's
 coefficients of
linear correlations, \cite{Pearson}.

\begin{figure}[ht]
 \noindent
\begin{center}
\epsfig{file=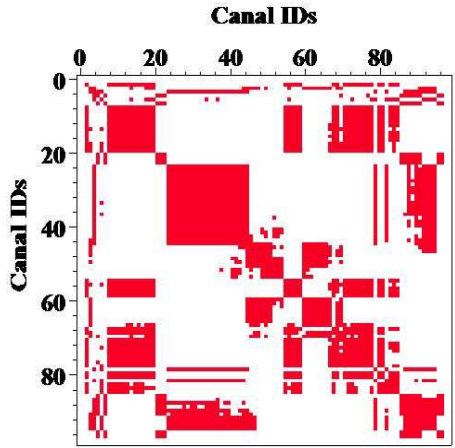, angle= 0,width =6cm, height =6cm}
  \end{center}
\caption{\small The coarse-grained
connectivity matrix
derived from  the
low-dimensional representation of  Venetian
canals given by the transition matrix ${\bf U}_6\,{\bf U}_6^\top $. }
\label{Fig2_08}
\end{figure}

The correlation matrix (\ref{Corr}) calculated with regard to the
first $n$  eigenvectors  possesses a complicated structure
containing the multiple overlapping blocks pertinent to a
low-dimensional
   representation of the network of
Venetian canals which allows
    for a further simplification.
In Fig.~\ref{Fig2_07}, we have presented
the correlation matrix (\ref{Corr}) figured out
for the first 7 eigenvectors of
the normalized Laplace operator
(\ref{norm_Lapalce})
 defined on the dual graph
representation of 96 Venetian canals.

Let ${\bf U}$ be the orthonormal matrix which contains
the eigenvectors $\{{\bf u}_k\}$, $k\,=\,1,\ldots,\, n-1,$
of the covariance (or correlation) matrix
as the row vectors. These vectors form the orthogonal
 basis of the $(n-1)$-dimensional vector space, in
which every variance $({\bf f}_k-{\bf m})$ is
represented by a  point ${\bf g}_k\,\in\,\mathbb{R}^{\,(n-1)}$,
\begin{equation}
\label{cov_transform}
{\bf g}_k\,=\,{\bf U}\,({\bf f}_k-{\bf m}\,).
\end{equation}
Then each original eigenvector
   ${\bf f}_k$ can be decoded from
${\bf g}_k\,\in\,\mathbb{R}^{\,(n-1)}$ by the inverse transformation,
\begin{equation}
\label{inverse_transff}
{\bf f}_k\,=\,{\bf U}^\top\,{\bf g}_k\,+\,{\bf m}.
\end{equation}
The use of transformations (\ref{cov_transform}) and (\ref{inverse_transff})
allows to obtain the $(n-1)$-dimensional representation
$\left\{\,\varphi_k \,\right\}_{\,k=1}^{\,(n-1)}$
of the $N$-dimensional basis vectors
$\left\{\, {\bf f}_s\,\right\}_{\,s=1}^{\,N}$
in the form
\begin{equation}
\label{compress}
\varphi_{\,k}\,=\,{\bf U}^\top{\bf U}\,{\bf f}_k \,+\,
\left(\,{\bf 1}-{\bf U}^\top{\bf U}\,\right)\,{\bf m},
\end{equation}
that minimizes the mean-square error between
 ${\bf f}_k\,\in\,\mathbb{R}^{\,N}$
and $\varphi_k\,\in\,\mathbb{R}^{\,(n-1)}$ for given $n$.

Variances of eigenvectors $\left\{\,{\bf f}_k\,\right\}$
are positively correlated within a principal component
 of the transport network. Thus, the
transition matrix ${\bf U}^{\top}{\bf U}$
can be interpreted as
the connectivity patterns acquired
by the network with respect to the diffusion process.
Two nodes, $i$ and $j$,
belong to one and the same principal
component  of the network
if $\left({\bf U}{\bf U}^\top\right)_{ij}\,>\,0$.
By applying the Heaviside function, which is
zero for negative argument and one for positive argument,
to the elements of the transition matrix
${\bf U}{\bf U}^\top$, we derive the
coarse-grained
connectivity matrix of network components.
In Fig.~\ref{Fig2_08}, we have shown the
coarse-grained
connectivity matrix
obtained from the
transition matrix ${\bf U}_6\,{\bf U}_6^\top $
for the dual graph representation of Venetian canals.

\subsection{Dynamical segmentations of transport networks}

In general, the building of low-dimensional
representations for transport networks
with respect to a certain dynamical process defined on them
 is a complicated procedure which cannot be reduced
 to (and reproduced by)
the naive
 introduction of
"supernodes" by either
 merging of several nodes or
shrinking complete subgraphs of the
original graph.
The implementation of spectral approach
 removes indeterminacies of the empirical clique
concatenation techniques
 used in space syntax analysis of urban textures, \cite{DaltonBafna}.
 If the covariance matrix clearly exhibits a block
structure, and once the relevant coarse-grained
connectivity matrix  is
computed, we can identify dynamical clusters (blocks) by using a
linearized cluster assignment and compute the cluster crossing,
the cluster overlap along the specified ordering using the
spectral ordering algorithm, \cite{Ding}.
\begin{figure}[ht]
 \noindent
\begin{center}
\epsfig{file=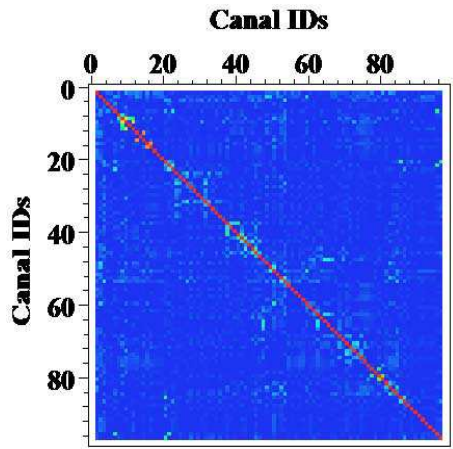, angle= 0,width =6cm, height =6cm}
  \end{center}
\caption{\small The covariance matrix calculated for the dual
 graph representation of Venetian canals with regrades
to the first 80 eigenvectors of the normalized Laplace
 operator (\ref{norm_Lapalce}). The entries of covariance matrix are
ranked from the largest positive values (red) to the
utmost negative values (blue).}
\label{Fig2_09}
\end{figure}
The problem of dynamical segmentations
 of a transport network
in fast time scales is more computationally complex
especially for large networks, because of
many eigenvectors if not all have to be taken
  into account while calculating the covariance matrix.
  It is important to note that the covariance
  matrix in this case takes the form of a sparse,  nearly
  diagonal matrix  (see Fig.~\ref{Fig2_09}).

\begin{figure}[ht]
 \noindent
\begin{center}
\epsfig{file=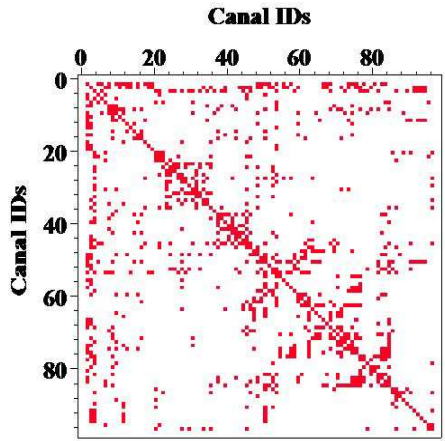, angle= 0,width =6cm, height =6cm}
  \end{center}
\caption{\small  The fast time scale
 coarse-grained  connectivity matrix for the
low-dimensional representation of    Venetian canals
  deduced from the transition matrix ${\bf U}_{79}\,{\bf U}_{79}^\top $.}
\label{Fig2_10}
\end{figure}
    Sparsity of the deduced coarse-grained
connectivity matrix (which is shown in  Fig.~\ref{Fig2_10})
  in fast time scales
entails  loosely coupled
  systems lack any form of large scale structure.
A sparse coarse-grained
connectivity matrix
 may be useful when storing and manipulating data
  for approximate descriptions of transport networks in fast time scales.

\begin{figure}[ht]
 \noindent
\begin{center}
\epsfig{file=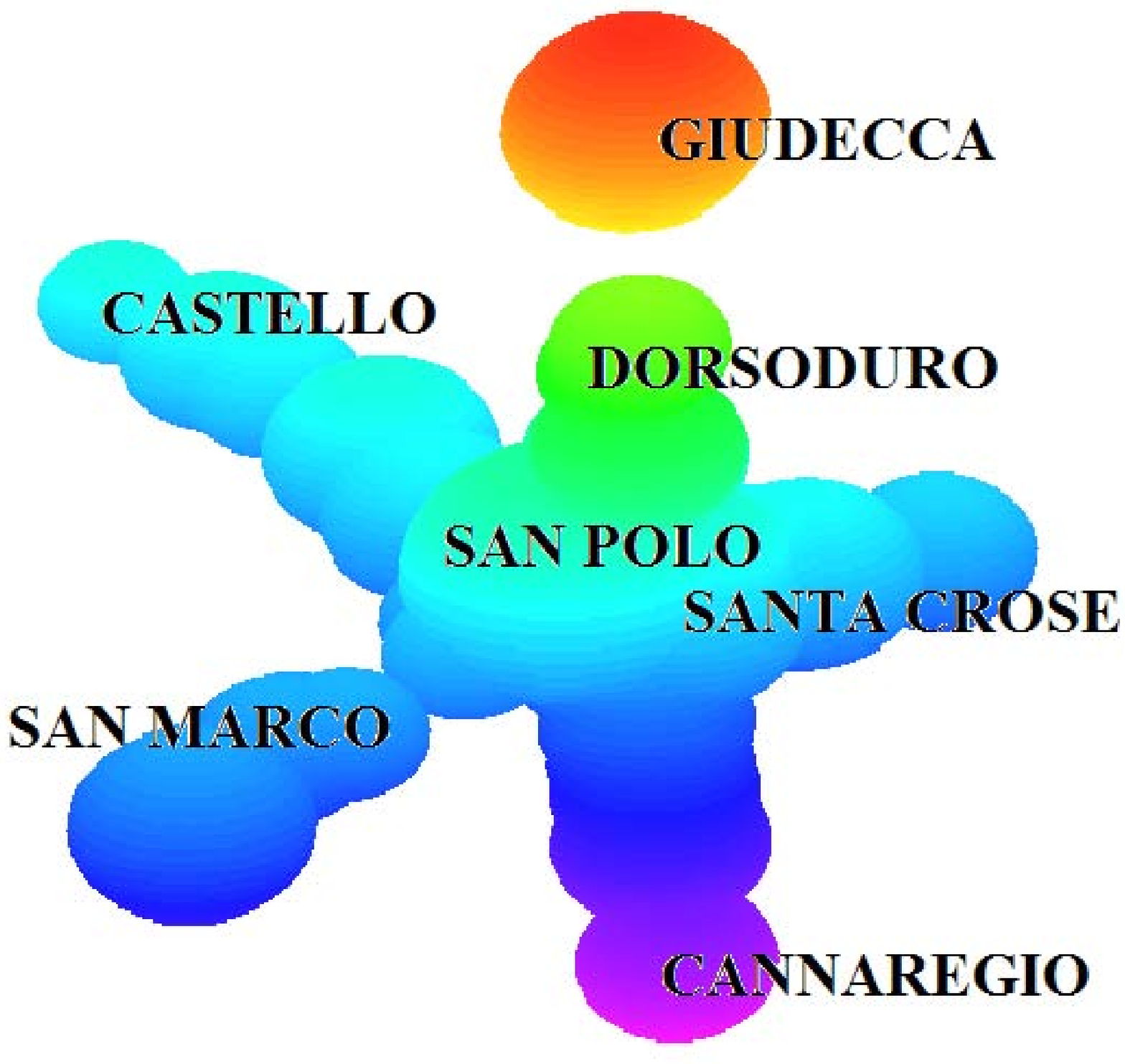, angle= 0,width =6cm, height =6cm}
  \end{center}
\caption{\small The 3D-image of  a
dynamical
segmentation of Venetian
 canals built for the first 8 eigenvalues of the normalized Laplace operator
(\ref{norm_Lapalce}). The structural differences between the
historical city districts of Venice are clearly visible.}
\label{Fig2_11}
\end{figure}

Low-dimensional representations of
not very large transport networks
given by
the coarse-grained connectivity matrices
can be represented by a 3D-graph.
 In Fig.~\ref{Fig2_11}, we  have shown
the 3D-image of  a
dynamical
segmentation of Venetian
 canals. The dual graph representation
of the Venetian canal network has been analyzed, and
a ball has been assigned to
each canal.
 The radius of the $i^\mathrm{th}$-ball is taken equal
to the norm (\ref{sqaured_norm}), the node $i$ has in the
$(N-1)$-dimensional Euclidean space associated
to random walks introduced on the connected, undirected
dual graph of Venetian canals,
\begin{equation}
\label{ball_rad}
r_i \, = \, \| \upsilon_i\|_{T}
\end{equation}
Those nodes
characterizing by the worst
accessibility levels
have the largest norms with respect to random walks
and therefore are represented by
balls of the largest radiuses.

The coordinates of each ball have been
 given by the relevant components
of the first three eigenvectors of the coarse-grained
connectivity matrix displayed in Fig.~\ref{Fig2_08}.
These eigenvectors determine
the directions of the largest variances
of correlations delineating
the low-dimensional representation of the network.
The key observation is that canals
with short access times
 are also characterized by small variances of correlations,
 therefore being forgathered
 proximate to the center of the
figure displayed
 in Fig.~\ref{Fig2_11}, no matter which city district they belong to.
In the contrary, the
worst accessible canals
are distinguished by
 the strongest correlation variances and
are
located on the figure fringes, far apart from
    its  center. At the same time, the radiuses of the balls
representing them are the largest among all other balls since they
acquire the utmost norms with respect to random walks.
It is remarkable that
they can be perfectly identified  with the
traditional historical sestieri of Venice.

\section{Discussion and Conclusion}
\label{Conclusion}
\noindent

The impact of urban landscapes on the construction of social
relations draws attention in the fields of ethnography, sociology,
and anthropology. In particular, it has been suggested that the
urban space combining social, economic, ideological and
technological factors is  responsible for the technological,
socioeconomic, and cultural  development, \cite{Low}. It is worth
to mention that the processes relating urbanization to economic
development and knowledge production are very general, being
shared by all cities belonging to the same urban system and
sustained across different nations and times \cite{Bettencourt}.
There is a tied connection between physical activity of humans,
their mobility  and the layout of buildings, roads, and other
structures that physically define a community \cite{Report2005}.
Spatial organization of a place has an extremely important effect
on the way people move through spaces and meet other people by
chance \cite{Hillerhanson}. The patterns of social movement and
economical development for a thousand years of Venetian history
have been imprinted in space syntax of the city.

In the present paper, we have investigated the canal network  in
Venice by means of random walks. Random walks being defined on an
undirected graph of $N$ modes, establish the $(N-1)$- dimensional
Euclidean space in which distances and angles acquire the clear
statistical interpretation. The properties of nodes with respect
to random walks allow partitioning the city canal network into
disjoint divisions which may be identified with the traditional
divisions of Venice (sestieri).

We have developed the general approach to the coarse-graining of transport networks
 based on the PCA method for the low-dimensional representation of large data set.
We believe that the proposed technique can be useful in many applications
 potential
applications such as the
establishing of rigorous mathematical
foundations for the analysis of
urban textures establishing the urbanization road to a harmonious  city.

\section{Acknowledgment}
\label{Acknowledgment}
\noindent

The work has been supported by the Volkswagen Foundation (Germany)
in the framework of the project: "Network formation rules, random
set graphs and generalized epidemic processes" (Contract no Az.:
I/82 418). The authors acknowledge the multiple fruitful
discussions with the participants of the workshop {\it Madeira
Math Encounters XXXIII}, August 2007, CCM - CENTRO DE CI\^{E}NCIAS
MATEM\'{A}TICAS, Funchal, Madeira (Portugal).

\end{document}